\documentclass[aps,prl,twocolumn,superscriptaddress]{revtex4-1}%
\usepackage[]{graphicx}
\usepackage{xcolor}%

\begin{document}
\title{Magnetic tuning of band topology evidenced by exotic quantum oscillations in the Dirac semimetal EuMnSb$_2$}
\author{Kan Zhao}
\email{Corresponding author: kan\_zhao@buaa.edu.cn}
\affiliation{Experimentalphysik VI, Center for Electronic Correlations and Magnetism, University of Augsburg, 86159 Augsburg, Germany}
\affiliation{School of Physics, Beihang University, Beijing 100191, China}
\author{Xuejiao Chen}
\thanks{These authors contributed equally to this work.}
\affiliation{CAS Key Laboratory of Magnetic Materials and Devices $\&$ Zhejiang Province Key Laboratory of Magnetic Materials and Application Technology, Ningbo Institute of Materials Technology and Engineering, Chinese Academy of Sciences, Ningbo 315201, China}
\author{Zhaosheng Wang}
\thanks{These authors contributed equally to this work.}
\affiliation{Anhui Province Key Laboratory of Condensed Matter Physics at Extreme Conditions, High Magnetic Field Laboratory of the Chinese Academy of Sciences, Hefei 230031, China}
\author{Jinyu Liu}
\affiliation{Department of Physics $\&$ Astronomy, University of California, Irvine, CA 92697, USA}
\author{Jiating Wu}
\author{Chuanying Xi}
\affiliation{Anhui Province Key Laboratory of Condensed Matter Physics at Extreme Conditions, High Magnetic Field Laboratory of the Chinese Academy of Sciences, Hefei 230031, China}
\author{Xiaodong Lv}
\affiliation{CAS Key Laboratory of Magnetic Materials and Devices $\&$ Zhejiang Province Key Laboratory of Magnetic Materials and Application Technology, Ningbo Institute of Materials Technology and Engineering, Chinese Academy of Sciences, Ningbo 315201, China}
\author{Lei Li}
\affiliation{Frontiers Science Center for Flexible Electronics, Xi'an Institute of Flexible Electronics (IFE) and Xi'an Institute of Biomedical Materials $\&$ Engineering, Northwestern Polytechnical University, 127 West Youyi Road, Xi'an 710072, China}
\author{Zhicheng Zhong}
\email{Corresponding author: zhong@nimte.ac.cn}
\affiliation{CAS Key Laboratory of Magnetic Materials and Devices $\&$ Zhejiang Province Key Laboratory of Magnetic Materials and Application Technology, Ningbo Institute of Materials Technology and Engineering, Chinese Academy of Sciences, Ningbo 315201, China}
\affiliation{College of Materials Science and Opto-Electronic Technology, University of Chinese Academy of Sciences, Beijing 100049, China}
\author{Philipp Gegenwart}
\email{Corresponding author: philipp.gegenwart@physik.uni-augsburg.de}
\affiliation{Experimentalphysik VI, Center for Electronic Correlations and Magnetism, University of Augsburg, 86159 Augsburg, Germany}

\begin{abstract}
Interplay between magnetism and electronic band topology is of central current interest in topological matter research. We use quantum oscillations as powerful tool to probe the evolution of band topology in the Dirac semimetal EuMnSb$_2$. The Eu local 4f magnetic moments display different antiferromagnetic states below 25~K and a field-polarized phase above 16~T. Upon cooling from 65 K into the field-polarized state, an exotic temperature dependent shift of oscillation peaks arises, accompanied by the development of non-zero Berry phase and a huge unconventional splitting of the oscillations. Band-structure calculations confirm the change from trivial to non-trivial band topology induced by the ferromagnetic Eu state, classifying EuMnSb$_2$ as unique magnetic topological semimetal.
\end{abstract}
\maketitle

Topological materials have received increasing attention in recent years due to their intriguing physical properties and potential applications~\cite{Na3Bi2012PRB, Cd3As22013PRB, Cd3As22014NM, Na3Bi2014Science, Cd3As2Crystal, Cd3As22014NC, Cd3As22015NM, WeylPRXDFT, WeylPRX, WeylScience, GrapheneNature, TIRMP}. Magnetism can introduce novel transport phenomena to topological systems. Quantum anomalous Hall effect has first been realized in the Cr- doped ferromagnetic (FM) topological insulator (Bi,Sb)$_2$Te$_3$~\cite{QAHE2013} and later in the intrinsic magnetic topological insulator MnBi$_2$Te$_4$~\cite{QAHE2020, QAHE2020NM}. Under time reversal symmetry breaking, Dirac semimetal will be converted into magnetic Weyl semimetal, with emergent magnetic monopole and chiral anomaly~\cite{WeylPRXDFT, WeylPRX, WeylScience, TIRMP}. Recently, the magnetic Weyl semimetal state has been realized in Kagome FM Co$_3$Sn$_2$S$_2$~\cite{Liu2018, Lei2018}.

The layered manganese pnictides AMn(Bi/Sb)$_2$ (A=Ca, Sr, Ba, Eu, and Yb) were reported to host Dirac or Weyl fermions, related to their two-dimensional (2D) Bi/Sb layers~\cite{Sr112PRL, Sr112PRB, Ca112PRB, Ca112APL, Sr112DFT, Sr112Neutron, Sr112APRES-PRB, Sr112APRES-SR, Eu112SA, BaMnSb2SR, Ba112PRB, Yb112PRB, SrMnSb2NM, Yb112ARPES, BaMnSb2PNAS, CaMnSb2PRB, Yb112caxis, YbMnSb2PRB, EuMnSb2017, Zhao2018, SrMnSb2018, CaNa2020}. The interplay between magnetism and Dirac dispersion induces novel quantum states: EuMnBi$_2$ and BaMnSb$_2$ show a bulk quantum Hall effect due to magnetically confined 2D Dirac fermions~\cite{Eu112SA, BaMnSb2020, BaMnSb2021}. YbMnBi$_2$ probably features a magnetic Weyl fermion state, due to spin degeneracy lifted by a canted FM component~\cite{Yb112ARPES}.

Splitting of the spin degeneracy of Dirac bands by magnetism has been claimed to be responsible for the non-trivial Berry phase of SrMnSb$_2$ with Mn vacancy~\cite{SrMnSb2NM}. However, the small net moment seems insufficient to achieve sizeable band splitting~\cite{SrMnSb2018, liuPRB2019, CPL2018, SrMnSb2019}. Taken into account the large moment of Eu$^{2+}$, the more promising system would be EuMn(Bi/Sb)$_2$.
The Shubnikov-de Haas (SdH) oscillation period in EuMnBi$_2$ is as low as 20 T~\cite{EuMnBi2018}, hindering a spin-splitting analysis, because the charge carriers will already approach the quantum limit, before Eu moments are fully polarized. This is different in EuMnSb$_2$, with oscillation period around 100 T~\cite{EuMnSb2021}.

In this Letter, we report exotic quantum oscillations in Dirac semimetal EuMnSb$_2$: an unusual temperature dependent shift of the high-field quantum oscillation pattern and spin splitting arises upon cooling to the field-polarized state of Eu moments. Berry phase analysis and accompanying band-structure calculations for the antiferromagnetic (AFM) low-field and FM high-field Eu states confirm the emergent non-trivial band topology.

\begin{figure}[t]
\includegraphics[width=0.48\textwidth]{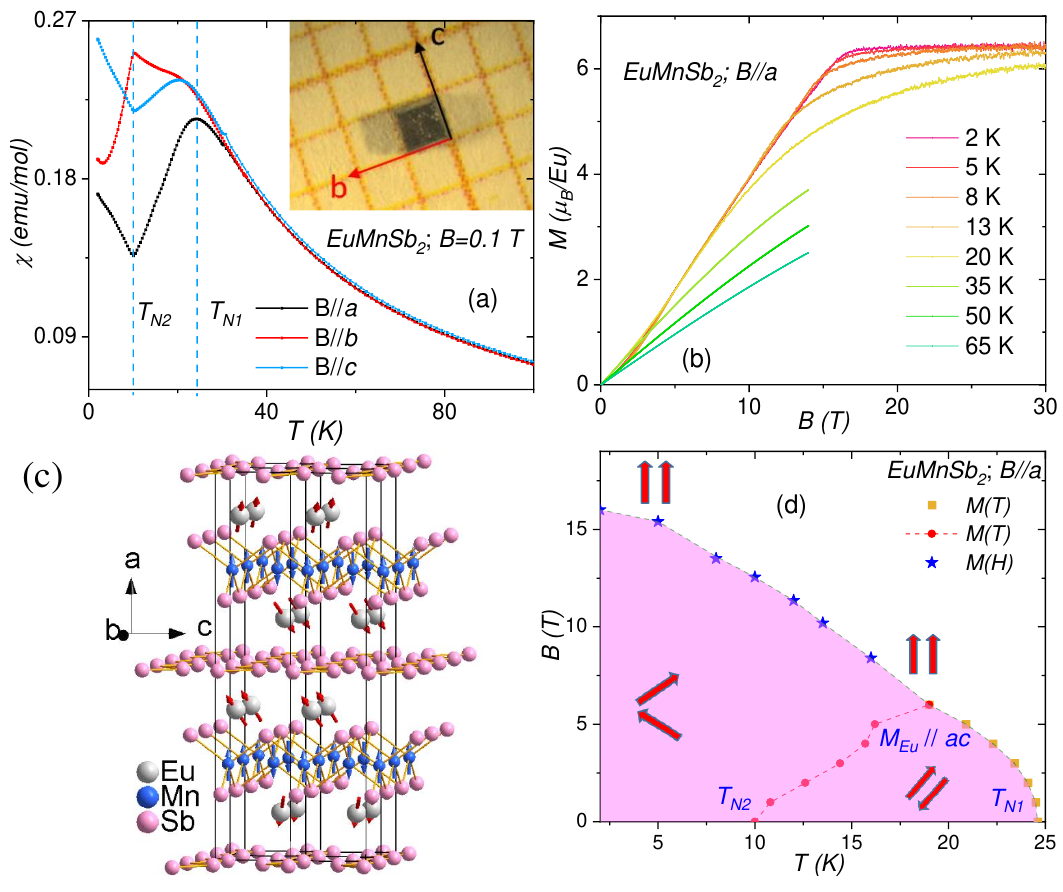}
\caption{\textbf{Magnetic structure and phase diagram of EuMnSb$_{2}$.} (a) Low-temperature susceptibility of EuMnSb$_{2}$ for B//$a$, B//$b$, and B//$c$ under 0.1 T, with an optical image of the single crystal as an inset. (b) Isothermal magnetization $M(H)$ of EuMnSb$_{2}$ for B//$a$ up to 30 T at various temperatures. (c) The crystal and magnetic structure of EuMnSb$_{2}$. Mn sublattices exhibit a C-type AFM order below 350 K. Eu sublattices exhibit an A-type AFM order below 10 K with 3D magnetic orders. (d) Phase diagram of EuMnSb$_{2}$ under B//$a$, including the magnetic susceptibility and magnetization data (see text).}
\label{Magnetic}
\end{figure}

We refer to supplemental material for details on synthesis, structural and magnetic characterization~\cite{Eu112SM}(see also references~\cite{PhysRevB.50.17953, perdew2008restoring, tran2009accurate, kresse1996efficient, stokes2005findsym, xu2020high, Luis2021, mostofi2008wannier90, wu2018wanniertools, kawamura2019fermisurfer, SdH2012} therein) and start with the magnetic properties of EuMnSb$_2$ shown in Fig.~\ref{Magnetic}. Above 30 K (Fig.~\ref{Magnetic}(a)), the susceptibility is isotropic along three crystal axes, consistent with the absent orbital component of Eu$^{2+}$ moments. Upon cooling, two subsequent phase transitions are resolved. Consistent with neutron scattering~\cite{EuMnSb2019, EuMnSb2020, EuMnSb2022}, Eu$^{2+}$ moments order in the canted A-type AFM structure below 24 K~\cite{EuMnSb2020, EuMnSb2021, EuMnSb20212}, which further rotates from ac plane towards b axis at $T_{N2}=10$ K~\cite{EuMnSb2022, Eu112SM}, clearly evidenced by the $\chi(T)$ dependence on along the three axes. Fig.~\ref{Magnetic}(c)~\cite{EuMnSb2022} shows the orientations of Mn and Eu moments below 10 K.

To map the phase diagram of Eu order for B//$a$, we conducted magnetic susceptibility in various fields, shown in Fig. S2(b)~\cite{Eu112SM}, and magnetization measurements up to 32 T at different temperatures, displayed in Fig. ~\ref{Magnetic}(b). Similar as in its sister compound EuMnBi$_2$~\cite{Eu112SA}, the Eu moment saturates at 6.5$\mu$$_{B}$ for 16 T at 2 K, slightly below 7$\mu$$_{B}$. Fig. ~\ref{Magnetic}(d) shows the derived phase diagram for B//$a$ with indicated Eu spin orientations.

\begin{figure*}[t]
\includegraphics[width=1.0\textwidth]{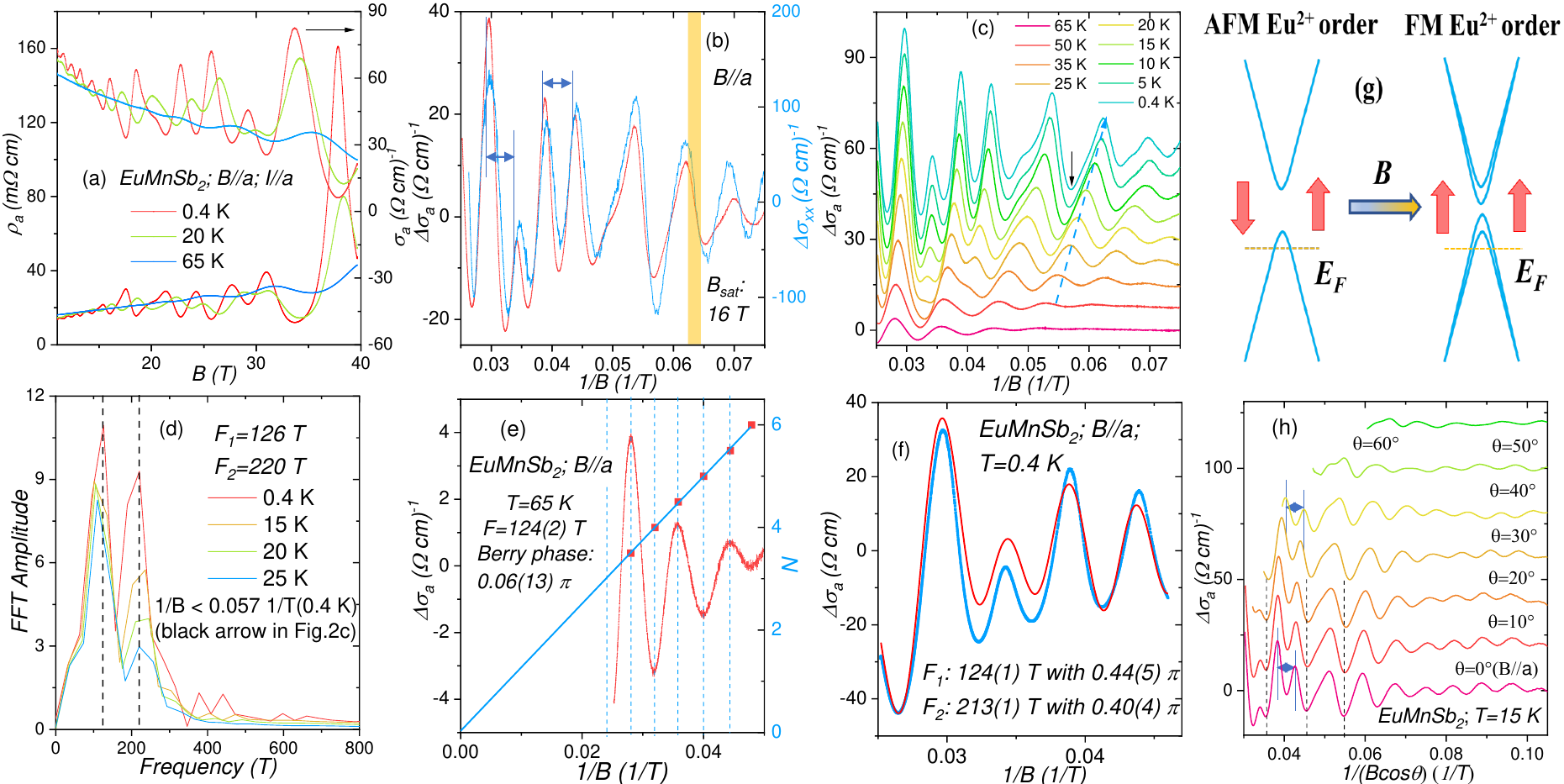}
\caption{\textbf{High field SdH investigation of EuMnSb$_{2}$.} (a) Out-of-plane resistivity ($\rho$$_a$) (left) and corresponding conductivity $\sigma$$_{a}$ (right) for B//a between 11 T and 39 T at 0.4 K, 20 K, and 65 K, respectively. (b) Quantum oscillations in the out of plane, $\sigma$$_{a}$ (0.4 K), and in plane conductivity, $\sigma$$_{xx}$ (5 K) vs. 1/B for B//a. (c) Quantum oscillations of $\sigma$$_{a}$ vs. 1/B for B//a up to 39 T at various temperatures, which are shifted for clarity. (d) FFT spectra of $\Delta$$\sigma$$_{a}$ for B//a under high field at various temperatures. (e) Landau level fan diagram of $\Delta$$\sigma$$_{a}$ at 65 K. The colored lines represent linear fits with integer LL indices assigned to the minimum of conductivity. (f) Quantum oscillations of $\Delta$$\sigma$$_{a}$ at 0.4 K with 1/B $< $ 0.046 T$^{-1}$, together with two band LK formula fitting curve in red color. (g) Qualitative illustration of the spin splitting of Dirac band from AFM to FM Eu order under magnetic field. (h) Angular dependence of high-field SdH oscillations at 15 K (see text).}
\label{SdH}
\end{figure*}

While SdH oscillations are visible already at low fields within the AFM regime and influenced by the Eu-moment rotation (cf. Fig. S5~\cite{Eu112SM}), we focus on the field-polarized state above 16 T in Fig.~\ref{SdH}.
Similar as (Sr/Ba)MnSb$_{2}$~\cite{BaMnSb2020, SrMnSb2NM}, EuMnSb$_{2}$ exhibits more pronounced SdH oscillations with current along the $a$ axis (Fig.~\ref{SdH}(a)), compared to the $bc$ plane case (Fig. S4(a)).
To clarify the Landau level (LL) structure, we analyze the SdH oscillations in the conductivity tensors, according to the formula $\sigma$$_{a}$= 1/$\rho$$_{a}$ and $\sigma$$_{xx}$ = $\rho$$_{xx}$/($\rho$$_{xx}$$^2$ +$\rho$$_{xy}$$^2$). Interestingly, the SdH oscillations of $\sigma$$_{xx}$ and $\sigma$$_{a}$ resemble each other, see Fig.~\ref{SdH}(b) (below 5 K) and Fig. S7(a) (15 K). Therefore, both SdH oscillations directly reflect the density of states in the LLs~\cite{Eu112SA}. 

Fig.~\ref{SdH}(c) presents the field dependence of magneto transport for EuMnSb$_{2}$ at various temperatures for B//$a$ up to 39 T. Generally, thermal fluctuations only lead to an oscillation damping but not to a shift of the peak values, as found for SrMnSb$_{2}$~\cite{SrMnSb2NM}. By contrast, as marked by the dashed non-vertical lines, oscillation peaks are not temperature independent, but significantly shift upon cooling. 
Similar as in EuMnBi$_{2}$~\cite{Eu112SA}, the Eu polarized moment in high fields B//$a$ continuously increases upon cooling from 65 K (Fig.~\ref{Magnetic}(c)), which seems closely related with the shift of the oscillation peaks.

In addition to the oscillation peak shift (cf. Fig.~\ref{SdH}(c), the huge splitting of the oscillations, indicated by the blue arrows in Fig.~\ref{SdH}(b) is most striking. The FFT analysis of the oscillations at $1/B <  0.057$ T$^{-1}$ is shown for different temperatures between 0.4 and 25 K in Fig.~\ref{SdH}(d). It displays the two frequencies 126 T and 220 T. This is ascribed to the splitting of the Dirac bands in the FM Eu state under high field. Note, that this splitting cannot be explained by the ordinary (Zeeman) spin splitting of quantum oscillations for the following reasons. (1) For the related SrMnSb$_2$~\cite{SrMnSb2NM} the Zeeman splitting is restricted to $T$ below 1.6 K and fields above 40 T. (2) As observed e.g. in ZrTe$_5$~\cite{ZrTe5-2016}, Zeeman energy $E$$_{Z}$= g${\mu}$$_{B}$B should gradually increase the oscillation splitting as lower LL are occupied in contrast to the observed equal distance of 0.005 T$^{-1}$, as marked by the two blue arrows in Fig.~\ref{SdH}(b). (3) The strong change in the ratio of peak amplitudes in high field and their significant temperature dependence speaks against the Zeeman origin of oscillation splitting.

At high $T$ the thermally disordered Eu moments cannot influence the electronic structure, and similarly as for SrMnSb$_2$~\cite{SrMnSb2018} a negligible Berry curvature is expected. Indeed, as shown in Fig. ~\ref{SdH}(e), the intercept on the LL index axis obtained from the extrapolation of the linear fit in the LL fan diagram is almost zero, i.e., 0.03(6) at 65 K, with the single oscillation frequency 124~T.
Fig.2(f) compares the 0.4 K oscillations in blue with a two-band Lifshitz-Kosevich fit in red~\cite{Eu112SM}, similar as applied in other topological semimetals~\cite{liu2019}.
The fit results frequencies of 124 and 213 T (that are almost identical to those from the FFT analysis) and corresponding Berry phases of 0.44(5) $\pi$ and 0.40(4) $\pi$, respectively. 
The left panel of Fig. ~\ref{SdH}(g) sketches the spin degenerate Dirac band in the low-field region at 15 K, with a single oscillation frequency of 126 T, as observed in Fig. S4(d-e). Upon further cooling, the FM Eu state induces band splitting as schematically shown in the right panel of Fig. ~\ref{SdH}(g), with two non-trivial oscillations at 124 and 213 T. 

\begin{figure*}[t]
\includegraphics[width=1.0\textwidth]{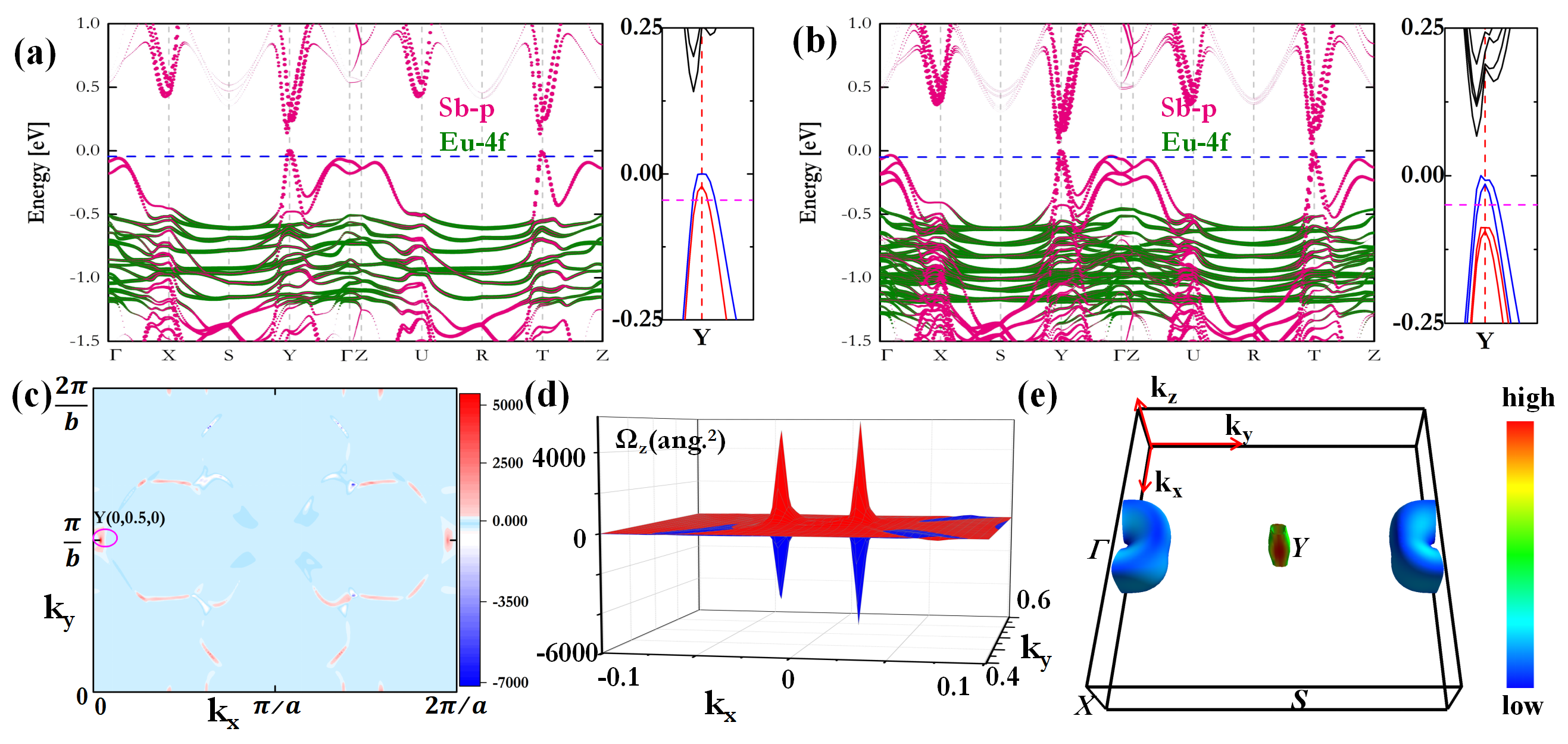}
\caption{\textbf{Electronic properties of EuMnSb$_{2}$.} The mBJ orbital projected band structures with Eu in (a) AFM-A and (b) FM spin configurations, with the spin configuration of Mn fixed in AFM-C type. Doping of 0.01 holes per unit cell reduces $E_F$ (indicated by dashed lines crossing the Dirac bands) by 0.045 eV (a) and 0.05 eV (b), respectively. (c) Color plot of Berry curvature distribution at the $k_z=0$ plane for the band structure of (b), calculated by GGA+U. (d) Berry curvature distribution near the Y-point (0,0.5,0) on the $k_z=0$ plane (first and second valence bands in red and blue, respectively). (e) Bulk Fermi surface of FM Eu state with $E_F= -50$ meV.}
\label{NodeLine}
\end{figure*}

To further study the origin of the unusual spin splitting, we measured the field-angle dependence of SdH oscillations when tilting the field from B//a towards the bc-plane. Fig.~\ref{SdH}(h) shows $\sigma$$_{a}$ plotted against 1/Bcos$\theta$ at 15 K in the FM Eu state. For the cylindrical quasi-2D Dirac bands in (Sr/Ba)MnSb$_2$, the overall behavior of the quantum oscillations plotted versus 1/Bcos$\theta$ is almost independent of $\theta$, except for a slight phase shift above 70$^{\circ}$~\cite{SrMnSb2NM, BaMnSb2020}. As shown in Fig.~\ref{SdH}(h), similar behavior has been observed for EuMnSb$_2$ only for $\theta${\textless}20$^{\circ}$. At larger angles, the oscillation pattern deviates from the vertical dashed line. 
This implies that the topology of Dirac bands is modified under tilted Eu moment orientation, leading to deviation from simple cylindrical quasi-2D Dirac bands in EuMnSb$_2$.

The normalized distance of oscillation peak at 15 K, indicated by arrows in Fig.~\ref{SdH}(h), has identical spacing (of 0.0044 T$^{-1}$) up to an angle of $\theta$=40$^{\circ}$. In case of g factor not exhibiting large variation with the angle, Zeeman effect scales with applied magnetic field so that the quantum oscillation splitting would be independent of angle. Thus, the angular dependent distance of oscillation peak should be attributed to the dominating spin splitting of Dirac bands in EuMnSb$_2$. 

To gain further inside, we performed DFT calculations of the electronic properties for the corresponding Eu$^{2+}$ magnetic configurations in EuMnSb$_2$.
The magnetic order of Mn is fixed in AFM-C type along the $a$ axis because of its strong exchange interaction, while the magnetic direction of Eu can be altered by external magnetic field and temperature.
For simplicity, the calculated magnetic ground state is AFM-A type of Eu along b axis. The mBJ band structure of EuMnSb$_2$ with this magnetic configurations is shown in Fig. \ref{NodeLine}(a)~\cite{Eu112SM}. There exists nearly linear Dirac bands from Sb-p orbital contributions around high symmetry points Y (0,0.5,0), as presented in the right side zoomed band structures of Fig. \ref{NodeLine}(a), where its band gap is about 0.16 eV. The double degeneracy of any k vector is realized by symmetry operator ($\bar{1}$|0,0,0)' that combines space inversion symmetry and time reversal symmetry.

The strong external magnetic field turns the magnetic configuration of Eu into the FM state (Fig. \ref{Magnetic}(d)).
Such a change of spin configuration causes the modification of electronic band structure because of the spin-textured band effect \cite{jiang2018spin,jiang2019stacking,liao2020materials,yang2021colossal}.
As shown in Fig. \ref{NodeLine}(b), four top doubly degenerate valence bands of Sb-p orbitals in AFM state are splitted in the FM case and the band gap decreases to 0.08 eV. Zoomed band structures around high symmetry points Y (right side of Fig. \ref{NodeLine}(b)) also clearly show the splitting of the degenerate linear bands.
Because of hole doping~\cite{Eu112SM}, realistic Fermi energy slightly crosses the two Dirac bands near Y with high Fermi velocity (as shown in Fig. \ref{NodeLine}(e)), in contrast to the low Fermi velocity of trivial bands at the $\Gamma$ point.
As shown in Fig.~S18, the two quantum oscillation frequencies observed at 0.4K under high field have been further verified by the DFT calculations~\cite{Eu112SM}. 

In addition, the touching node line along the high symmetry path T to Z (0.5,0.5,0.5) is still kept in the FM case. It is attributed to non-symmorphic glide plane (m$_{y}$|0,0,$\frac{1}{2}$)' generating the two-dimensional irreducible representation G$_2$G$_2$(2) (as shown in Fig. \ref{NodeLine}(b) and Fig. S11). Here m$_{y}$ shows mirror at k$_y$=0, (0,0,$\frac{1}{2}$) means translated half distance along out of plane and time reversal symmetry combines with these two operators.
Furthermore, we construct an enclosed k path around the T point to calculate the Wilson loop which shows a $\pi$ Berry phase (Fig. S12 and S13)~\cite{sun2017dirac}.
Here, the quantized non-trivial geometrical phase of occupied electrons might make a contribution to the quantum oscillation signal.
Moreover, we calculated the Berry curvature located at k$_z$=0 for the whole plane and enlarge the region around the Y point (0,0.5,0) for two occupied states as illustrated in Fig. \ref{NodeLine}(c,d).
It can be seen that there are Berry curvature peaks near this Y-point.
Two Berry curvature peaks' symbols for the specific occupied states (red or blue of Fig. \ref{NodeLine}(d)) are the same, which might be similar to the two same chiral Weyl points with a small gap.
Owing to occupied top linear band electrons contributing to the transport signal, the peak of Berry curvature around the Y point and the nodal line along the T to Z path provides important theoretical insight to the anomalous experimental results.

The complex magnetic phase diagram with different configurations of Eu$^{2+}$ localized 4f magnetic moments is key to understand the anomalous quantum oscillations in EuMnSb$_2$. The observed spin splitting of degenerate Sb-p orbital Dirac bands and emergence of non-trivial Berry curvature, indicated by the exotic temperature dependence of the quantum oscillations, opens an interesting perspective to design and control novel topological phases by the help of local 4f electrons.

In conclusion, we observe anomalous magneto quantum transport with a modification of SdH oscillation pattern with temperature in the magnetic topological semimetal EuMnSb$_2$. 
Our results comprehensively demonstrate the entire process how magnetism modifies the Dirac band with spin degeneracy to acquire a non-zero Berry phase. EuMnSb$_2$ semimetal in large magnetic field exemplifies the intriguing quantum transport behavior arising in magnetic topological materials.

\acknowledgments
The authors would like to thank Igor Mazin, Yurii Skourski, Zhiming Wang, Zhaoliang Liao, Zhigao Sheng, Johannes Knolle, and Valentin Leeb for helpful discussions and experimental support.

This work was partially supported by the Science Center of the National Science Foundation of China (52088101), the National Key R\&D Program of China (Grant No. 2021YFA0718900
and No. 2017YFA0303602), the Key Research Program of Frontier Sciences of CAS (Grant No. ZDBS-LY-SLH008), the National Nature Science Foundation of China (Grants
No. 11974365 and No. 51931011), K.C. Wong Education Foundation (GJTD-2020-11).
The work in Augsburg was supported by the German Science Foundation through SPP1666 (project no. 220179758), TRR80 (project no. 107745057) and via the Sino-German Cooperation Group on Emergent Correlated Matter. The work in Beijing was supported by the National Natural Science Foundation of China (Grants No. 12274015), the Beijing Nova Program (Grant No. Z211100002121095), and the Fundamental Research Funds for the Central Universities. The work in Hefei was supported by the National Natural Science Foundation of China (Grants No. 11874359). A portion of this work was performed on the Steady High Magnetic Field Facilities, High Magnetic Field Laboratory, Chinese Academy of Sciences, and supported by the High Magnetic Field Laboratory of Anhui Province.



\bibliography{Eu112}

\end{document}